\documentclass[prd,amsmath,amssymb,preprintnumbers,superscriptaddress,twocolumn,10pt,nofootinbib]{revtex4-1} 

\pdfoutput=1
\usepackage{graphicx}
\usepackage{subfigure}
\usepackage{float}

\usepackage{dcolumn}
\usepackage{bm}
\usepackage{amssymb}
\usepackage{latexsym}
\usepackage{booktabs}
\usepackage{amsmath}
\usepackage{multirow}
\usepackage{url}
\usepackage{changes}
\usepackage{soul} 
\usepackage{color, xcolor}
\usepackage[colorlinks=true, linkcolor=red, citecolor=blue]{hyperref}

\hypersetup{
    colorlinks=true,
    linkcolor=red,
    citecolor=blue
}
\IfFileExists{orcidlink.sty}{\usepackage{orcidlink}}{%
  \newcommand{\orcidlink}[1]{%
    \href{https://orcid.org/##1}{%
      \includegraphics[height=1.6ex]{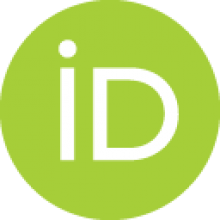}%
    }%
  }%
}

\usepackage[normalem]{ulem}
\usepackage{array}
\usepackage{enumerate}

\def\be{\begin{equation}}
\def\ee{\end{equation}}

\def\bea{\begin{eqnarray}}
\def\eea{\end{eqnarray}}
\newcommand{\bes}{\begin{equation*}}
\newcommand{\ees}{\end{equation*}}
\newcommand{\beqa}{\begin{eqnarray}}
\newcommand{\eeqa}{\end{eqnarray}}

\begin{document}

\title{Synergy between CSST and future gravitational-wave detectors: Probing primordial black holes by cross-correlating dark sirens with galaxies}

\author{Ya-Nan Du\orcidlink{0009-0003-0453-9046}}
\affiliation{Liaoning Key Laboratory of Cosmology and Astrophysics, College of Sciences, Northeastern University, Shenyang 110819, China}

\author{Ji-Yu Song\orcidlink{0009-0003-8111-0470}}
\affiliation{Liaoning Key Laboratory of Cosmology and Astrophysics, College of Sciences, Northeastern University, Shenyang 110819, China}

\author{Jing-Fei Zhang\orcidlink{0000-0002-3512-2804}}
\affiliation{Liaoning Key Laboratory of Cosmology and Astrophysics, College of Sciences, Northeastern University, Shenyang 110819, China}

\author{Xin Zhang\orcidlink{0000-0002-6029-1933}}\thanks{Corresponding author: zhangxin@neu.edu.cn}
\affiliation{Liaoning Key Laboratory of Cosmology and Astrophysics, College of Sciences, Northeastern University, Shenyang 110819, China}
\affiliation{MOE Key Laboratory of Data Analytics and Optimization for Smart Industry, Northeastern University, Shenyang 110819, China}
\affiliation{National Frontiers Science Center for Industrial Intelligence and Systems Optimization, Northeastern University, Shenyang 110819, China}

\begin{abstract}
Gravitational-wave (GW) events and galaxies both trace the cosmic matter distribution, but the mergers of astrophysical black holes and primordial black holes (PBHs) are expected to populate different environments and therefore to cluster with different biases. The GW clustering bias is thus a statistical observable that can separate the two populations. We assess how well this can be done by cross-correlating the photometric galaxy survey of the Chinese Space-station Survey Telescope (CSST) with mock GW catalogs from two future detector networks: the third-generation ET2CE network (the Einstein Telescope and two Cosmic Explorer detectors) and the multiband BDET2CE network, which adds the space-based baseline Decihertz Interferometer Gravitational-Wave Observatory. We find that CSST combined with 10 years of ET2CE observations can reveal a PBH contribution once its fraction in the local total merger rate exceeds about $40\%$, while the much sharper sky localization of BDET2CE lowers this threshold to about $20\%$. The improvement comes from recovering the small-scale clustering information that localization errors would otherwise erase. These results show that combining future GW detector networks with CSST galaxy clustering offers a promising and largely independent route to identifying PBHs statistically.
\end{abstract}



\maketitle

\section{Introduction}\label{sec:Introduction}

Since the LIGO-Virgo Collaboration reported the first direct detection of gravitational waves (GWs) \cite{LIGOScientific:2016aoc}, a growing catalog of GW events has sharpened our view of the binary black hole (BBH) population \cite{LIGOScientific:2025pvj}. The black holes in these mergers may have two distinct origins: astrophysical black holes (ABHs) formed through stellar evolution \cite{Barack:2018yly}, and primordial black holes (PBHs) formed in the early universe when enhanced curvature perturbations reentered the horizon and collapsed \cite{Carr:1974nx, Carr:2016drx, Carr:2021bzv, Zhang:2023tfv, Zhang:2023zmb, Domenech:2026nun, Pi:2024ert}. Such perturbations may be seeded by non-Gaussianity \cite{DeLuca:2019qsy, Yoo:2018kvb}, inflationary dynamics \cite{Choudhury:2013woa, Zhou:2020kkf, Pi:2021dft}, or phase transitions \cite{Hawking:1982ga, Moss:1984zf}, allowing PBHs to span a wide range of masses \cite{Sasaki:2018dmp, Bellomo:2017zsr, Jia:2026psi, Jia:2025vqn, Yang:2024vij, Yang:2025uvf, Yang:2024pfb, LHAASO:2025kyn, Zhao:2025ekx, Zhao:2025ddy, Zhao:2024jad, Hong:2026rcl, Fumagalli:2024kgg, Wang:2021djr, Calza:2024fzo, Calza:2024xdh, Calza:2025mwn}. Stellar-mass PBHs are especially intriguing because their masses overlap with the BBH events observed by the LIGO-Virgo-KAGRA (LVK) Collaboration \cite{Bird:2016dcv, DeLuca:2020qqa, Franciolini:2021tla, Andres-Carcasona:2026avd}. However, the waveform of a single merger does not encode its formation channel, so ABH and PBH origins can be separated only statistically, through population-level signatures. Developing robust methods to identify a PBH contribution therefore remains a central challenge in GW cosmology.

A promising population-level signature is the spatial clustering of the merger events. As products of stellar evolution, ABHs reside mainly in galaxies hosted by massive dark matter halos, and their clustering bias grows with redshift \cite{Peron:2023zae}. PBHs, by contrast, are themselves a dark matter candidate: binaries formed in the early universe trace the dark matter distribution directly, while those assembled at late times occur preferentially in low-mass halos; in both cases the mergers are more uniformly distributed and carry a lower bias that evolves only weakly with redshift \cite{Raccanelli:2016cud, Holst:2024ubt}. The clustering bias of GW sources, and in particular its redshift evolution, therefore offers a statistical handle for separating the two populations. This idea rests on the established picture of GW events as tracers of the large-scale structure (LSS) of the Universe \cite{Namikawa:2015prh, Bird:2016dcv, Diaz:2021pem}: future detectors with wide sky coverage and precise luminosity-distance measurements can constrain the host-halo bias of GW sources in luminosity-distance space, even when individual sky localizations are modest \cite{Libanore:2020fim, Libanore:2021jqv}. Building on this, Ref.~\cite{Raccanelli:2016cud} first proposed using the bias as a discriminator between ABH and PBH binaries.

GW standard sirens have become a powerful and independent probe of the cosmic expansion history. A small fraction of events are bright sirens, whose electromagnetic counterparts pinpoint the host-galaxy redshift \cite{Schutz:1986gp, Zhao:2010sz, Cai:2016sby, Wang:2018lun, Zhang:2018byx, Zhang:2019ylr, Zhang:2019loq, Wang:2019tto, Jin:2020hmc, Qi:2021iic, Jin:2021pcv, Wang:2021srv, Wu:2022dgy, Jin:2022tdf, Wang:2022oou, Jin:2023tou, Yu:2023ico, Jin:2023sfc, Han:2023exn, Jin:2023zhi, Feng:2024mfx, Feng:2024lzh, Han:2024sxm, Han:2025fii, Feng:2025wbz, Zhao:2019gyk}, while the vast majority are dark sirens that provide a luminosity distance but no direct redshift. For dark sirens, the redshift is commonly recovered by statistically associating each event with candidate host galaxies in a catalog, together with a population model of the sources \cite{Jin:2025dvf, Chen:2017rfc, Gray:2019ksv, Yu:2020vyy, Zhu:2021aat, Zhu:2021bpp, Jin:2022qnj, Li:2023gtu, Yun:2023ygz, Xiao:2024nmi, Zhu:2024qpp, Zheng:2024mbo, Song:2025ddm, Song:2025bio, Zhang:2025yhi, Dong:2025ikq, Dong:2026uxr, Xiao:2025mcg, Xiong:2026piv}. A complementary and fully statistical route, introduced by Ref.~\cite{Oguri:2016dgk}, is to cross-correlate GW sources with galaxies: assuming that GW sources reside in galaxies, the galaxy survey supplies redshifts and angular positions while GW observations supply luminosity distances, and the two tracers correlate only through the distance-redshift mapping \cite{Du:2025odq, Nair:2018ign, Mukherjee:2019wcg, Bera:2020jhx, Diaz:2021pem, Ghosh:2023ksl}. Although this cross-correlation is most often used to constrain the expansion history, here we turn it into a probe of the GW clustering bias: with the background cosmology fixed, the amplitude of the cross-spectrum is controlled by the GW bias, giving direct access to the population-level signature that distinguishes PBHs from ABHs. Relative to event-by-event host searches \cite{Song:2022siz, Song:2026kii}, this statistical approach is less sensitive to incomplete galaxy catalogs and population-model assumptions, and directly measures the clustering of GW sources. Such measurements are, however, beyond present capabilities: current LVK catalogs remain limited in event number, redshift reach, and sky localization \cite{LIGOScientific:2025pvj}, so the method calls for the next generation of GW detectors together with deep, wide galaxy surveys.

Realizing this program requires two ingredients: a GW detector network that delivers a large, well-localized sample of events, and a galaxy survey that is both wide and deep. On the GW side, the third-generation ground-based detectors, the Einstein Telescope (ET) \cite{ET:2025xjr, Punturo:2010zz} and the Cosmic Explorer (CE) \cite{LIGOScientific:2016wof}, will detect large numbers of BBH mergers with substantially improved sky localization \cite{Song:2022siz}; adding the space-based baseline Decihertz Interferometer Gravitational-Wave Observatory (B-DECIGO) \cite{Nakamura:2016emq} broadens the frequency coverage and sharpens the localization of individual events by a further two to three orders of magnitude \cite{Dong:2024bvw}. We consider two representative configurations: ET2CE, combining ET with two CE detectors, and BDET2CE, which further includes B-DECIGO. Because the sky-localization accuracy of GW sources sets how much of the cross-correlation signal survives, this improvement is decisive for measuring the GW clustering bias \cite{Song:2026tdy}. On the galaxy side, ongoing and upcoming surveys, including the Dark Energy Spectroscopic Instrument \cite{DESI:2016fyo}, Euclid \cite{EUCLID:2011zbd, Euclid:2024yrr}, the Vera C. Rubin Observatory \cite{LSSTScience:2009jmu}, and the Chinese Space-station Survey Telescope (CSST) \cite{CSST:2025ssq}, will map large galaxy samples over wide areas and broad redshift ranges, providing powerful probes of LSS \cite{Bonvin:2011bg, DiDio:2013sea, Kaiser:1987qv, Szalay:1997cc, Yoo:2009au}. Among them, CSST is especially well suited to this work: its wide and deep photometric survey yields a dense galaxy sample reaching the high redshifts where the PBH contribution to the merger rate is most prominent, supplying exactly the redshift and clustering information that GW data alone cannot provide.

In a previous study, we forecast the cross-correlation of CSST galaxies with GW dark sirens detected by third-generation ground-based detectors, and showed that it can constrain cosmological parameters and the GW clustering bias \cite{Du:2025odq}. Here we build on that framework but shift the goal from measuring cosmology to identifying a PBH contribution, using the GW clustering bias as the discriminating observable. We model the merger-rate and redshift distributions of the ABH and PBH populations, adopting an early-PBH merger-rate model, and simulate mixed BBH catalogs observed by ET2CE and BDET2CE. Combining these simulated GW samples with a mock CSST photometric galaxy catalog, we compute the tomographic galaxy autocorrelation, galaxy-GW cross-correlation, and GW autocorrelation angular power spectra, and use the Fisher information matrix (FIM) to forecast the precision on the galaxy and GW bias parameters and on the PBH fraction. We find that the cross-correlation substantially improves the sensitivity to a PBH contribution over GW data alone, and that the superior localization of the multiband network lowers the PBH fraction at which such a contribution can be identified.

This paper is organized as follows. Section~\ref{sec:method} describes the method, Sec.~\ref{sec:results} presents the results and discussion, and Sec.~\ref{sec:conclusions} summarizes our conclusions.

\section{Methods}\label{sec:method}

In this work, we use angular power spectra as the main statistical tool to quantify the clustering of galaxy surveys and GW sources as tracers of cosmic LSS. We first simulate a GW event catalog containing both ABH and PBH merger events to obtain the number density for a given detector configuration. In the subsequent angular power-spectrum analysis, the simulated catalog is mainly used to determine the GW number density and to assess the effects of luminosity-distance and sky-localization uncertainties, while the specific sky positions of individual events do not directly enter the angular power-spectrum calculation. Meanwhile, we model the redshift distribution of the galaxy survey. The clustering signals of both GW sources and galaxies are described through their bias factors and the underlying matter density fluctuations in the theoretical angular power-spectrum model. We then compute the GW-galaxy cross-correlation, galaxy autocorrelation, and GW autocorrelation angular power spectra, and estimate the parameter uncertainties with the FIM method. Throughout this analysis, we fix the background cosmology to the Planck 2018 $\Lambda$ cold dark matter ($\Lambda$CDM) model, with $h = 0.6766$, $\Omega_{\rm c} = 0.26069$, $\Omega_{\rm b} = 0.04897$, $n_{\rm s} = 0.9665$, and $A_{\rm s} = 2.1\times10^{-9}$.

\subsection{Simulating ABH and PBH source GW catalogs}\label{ssec:GW catalogs}

To construct a simulated GW catalog containing both ABH and PBH merger events, we model their redshift distributions. We assume that the BBH merger rate follows the redshift evolution of the star formation rate described by the Madau-Dickinson model and adopt the following parametric form:
\begin{equation}\label{eq:1}
\mathcal{R}(z) = \mathcal{R}_0^{\mathrm{tot}} \frac{(1+z)^\gamma}{1 + \left[(1+z)/(1+z_{\mathrm{p}})\right]^{\gamma+\kappa}},
\end{equation}
where $\mathcal{R}_0^{\mathrm{tot}}$ is the total merger rate at $z=0$, set to $\mathcal{R}_0^{\mathrm{tot}} = 27\,\mathrm{Gpc}^{-3}\,\mathrm{yr}^{-1}$. For the ABH population, we choose $z_{\mathrm{p}} = 2$, $\gamma = 2.7$, and $\kappa = 3$ \cite{Libanore:2023ovr}, which gives
\begin{equation}\label{eq:2}
\mathcal{R}^{\mathrm{A}}(z) = \mathcal{R}_0^{\mathrm{A}} \frac{(1+z)^{2.7}}{1 + \left[(1+z)/3\right]^{5.7}}.
\end{equation}
For the PBH population, we adopt the early-merger-rate model with $\gamma = 1.17$ and $\kappa = -1.17$ \cite{Berti:2025usa}, which reduces the merger-rate evolution to a simple power law:
\begin{equation}\label{eq:3}
\mathcal{R}^{\mathrm{P}}(z) = \mathcal{R}_0^{\mathrm{P}} (1+z)^{1.17},
\end{equation}
where the parameter $z_{\mathrm{p}}$ is omitted.

To characterize the PBH contribution to the total merger rate, we introduce the fraction parameter $f^{\mathrm{P}}$, defined as the ratio of the PBH merger rate at $z=0$ to the total merger rate,
\begin{equation}\label{eq:4}
\mathcal{R}_0^{\mathrm{P}} = f^{\mathrm{P}} \mathcal{R}_0^{\mathrm{tot}}, \quad \mathcal{R}_0^{\mathrm{A}} = (1 - f^{\mathrm{P}}) \mathcal{R}_0^{\mathrm{tot}}.
\end{equation}
With this formulation, the redshift evolution of the total merger rate can be expressed as,
\begin{equation}\label{eq:5}
\begin{aligned}
\mathcal{R}^{\mathrm{tot}}(z) =&  \mathcal{R}^{\mathrm{A}}(z) + \mathcal{R}^{\mathrm{P}}(z) \\
                              =&  \mathcal{R}_0^{\mathrm{tot}} (1-f^{\mathrm{P}}) \frac{(1+z)^{2.7}}{1 + \left[(1+z)/3\right]^{5.7}} \\
                              & + \mathcal{R}_0^{\mathrm{tot}}  f^{\mathrm{P}} (1+z)^{1.17}.
\end{aligned}
\end{equation}

We employ the Power Law + Peak model \cite{Talbot:2018cva} for the ABH mass distribution. For the PBH population, we follow Ref.~\cite{Ng:2022agi} and adopt a lognormal mass distribution. Because the masses of both populations are assumed to lie in the range $5$--$100\,M_{\odot}$, their merger waveforms are intrinsically indistinguishable \cite{Libanore:2023ovr}. The redshift evolution of the GW clustering bias therefore provides a statistical way to distinguish between the ABH and PBH merger populations.

In terms of bias modeling, we parametrize the bias of the ABH population as,
\begin{equation}\label{eq:6}
b^{\mathrm{A}}(z)=C(1+z)^D,
\end{equation}
where $C = 1.2$ and $D = 0.59$ \cite{Peron:2023zae}. Because we focus on early PBH mergers, and under the common approximation that early PBHs Poisson-sample the underlying dark matter field on LSS, their clustering bias is expected to be close to unity. We therefore adopt the simplified constant bias model $b^{\mathrm{P}} = 1$, following Ref.~\cite{Libanore:2023ovr}. Since the detectability of PBHs depends on the choice of $b^{\mathrm{P}}$, we quantify in Section~\ref{ssec:GW bias} how this parameter affects the detection threshold of $f^{\mathrm{P}}$. Since the survey cannot identify the origin of each merger progenitor, the observed effective black hole merger bias is computed by weighting the ABH and PBH biases by their relative merger rates at each redshift,
\begin{equation}\label{eq:7}
b_{\mathrm{eff}}(z) = \frac{\mathcal{R}^{\mathrm{A}}(z)}{\mathcal{R}^{\mathrm{tot}}(z)} b^{\mathrm{A}}(z) + \frac{\mathcal{R}^{\mathrm{P}}(z)}{\mathcal{R}^{\mathrm{tot}}(z)} b^{\mathrm{P}}(z).
\end{equation}

We denote the model containing both ABH and PBH contributions as the AP model, and the model containing only ABH mergers as the A model. The A model serves as the reference model for assessing whether a PBH contribution can be distinguished from the pure-ABH scenario. In contrast, the AP model incorporates the effect of $f^{\mathrm{P}}$ on the redshift evolution of the effective GW bias through Eq.~\eqref{eq:7}, enabling direct constraints on $f^{\mathrm{P}}$.

\begin{figure*}
    \centering
    \includegraphics[width=0.464\linewidth]{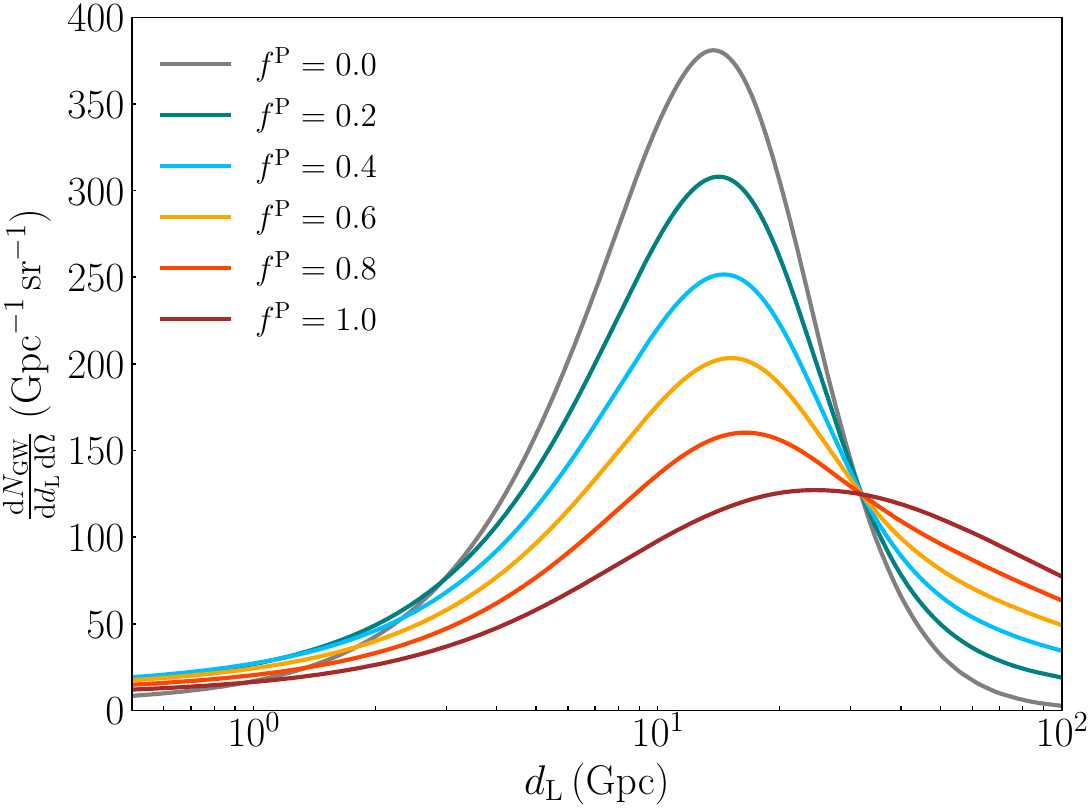}
    \hspace{0.5cm}
    \includegraphics[width=0.45\linewidth]{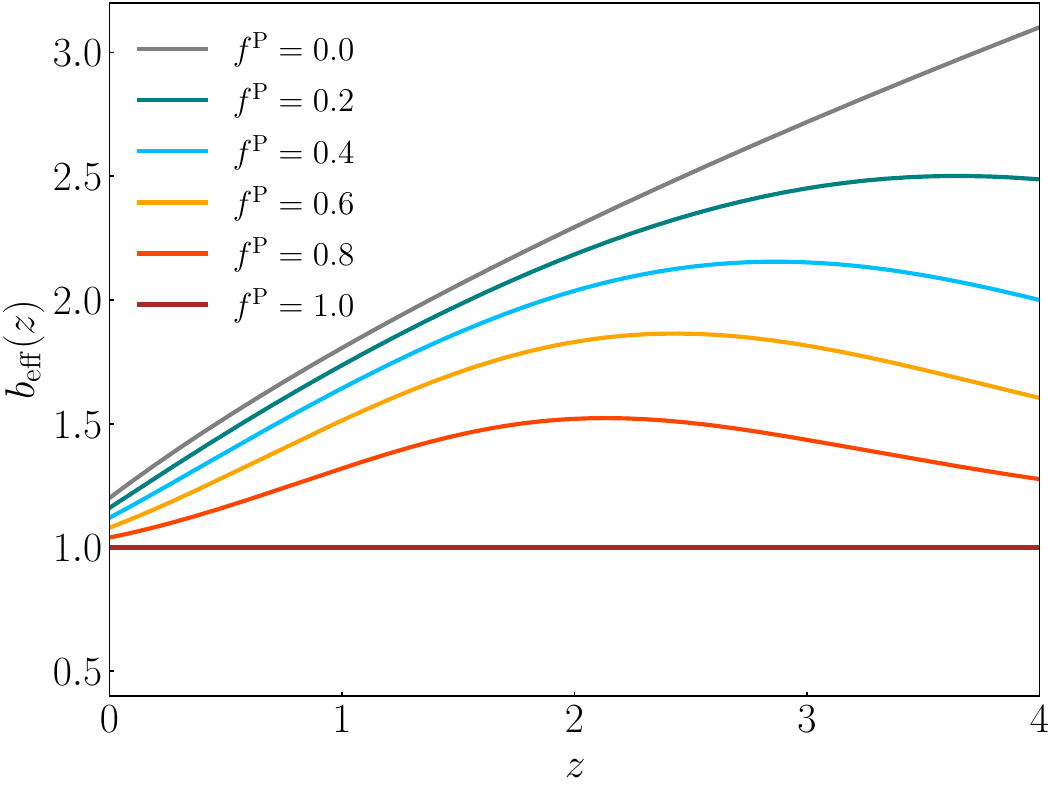}
    \caption{Detected GW number density and effective GW bias for different PBH fractions. The left panel shows the number density of detected GW sources as a function of luminosity distance for one year of observation, while the right panel shows the redshift evolution of the effective GW bias. Larger $f^{\mathrm{P}}$ values increase the high-redshift event density and flatten or lower the effective bias.}
    \label{fig:GW_num_and_b_eff}
\end{figure*}

Figure~\ref{fig:GW_num_and_b_eff} shows the number density distribution of detected GW sources (left) and the evolution of the effective GW bias with redshift (right) over a one-year observation period. Here $f^{\mathrm{P}} = 0$ corresponds to a pure ABH population. The left panel shows that increasing $f^{\mathrm{P}}$ raises the number density of GW events at large luminosity distances, corresponding to high redshifts. The right panel shows that the effective bias becomes flatter or decreases at high redshift as $f^{\mathrm{P}}$ increases. This redshift dependence provides an observational signature for identifying a PBH contribution through clustering.

To predict the measurement accuracy of the ET2CE detector network and the BDET2CE multiband joint detection network, we simulate the instrumental uncertainties of GW events using the publicly available code \textsc{GWFish}\footnote{\url{https://github.com/janosch314/GWFish}} \cite{Dupletsa:2022scg}. Following the selection criteria in Ref.~\cite{Du:2025odq}, we retain events with signal-to-noise ratio $\rho>\rho_{\rm th}=8$ and relative luminosity-distance error $\Delta d_{\mathrm{L}}/d_{\mathrm{L}}<1$ as the final effective GW sample.

\subsection{Modeling galaxy catalogs}\label{ssec:galaxy catalogs}

For the galaxy survey model, we use the photometric survey catalog expected from CSST. The catalog covers $0<z<4$ over an area of approximately $17500\,\mathrm{deg}^2$. We adopt the galaxy number density distribution from Ref.~\cite{Gong:2019yxt}, which contains about $1.925\times10^9$ galaxies and includes a photometric redshift uncertainty of $0.05(1+z)$. We divide the redshift interval $0<z<4$ into 15 tomographic bins containing equal numbers of galaxies. When computing the angular power spectra, we assume that the galaxy distribution is statistically isotropic over the observed sky and include the sky coverage fraction $f_{\mathrm{sky}}$ to account for the CSST survey footprint.

For the CSST galaxy clustering bias, we adopt the parametrization
\begin{equation}\label{eq:8}
b_{\mathrm{g}}(z) = b_{0}(1+z)^{b_{1}},
\end{equation}
with $b_{0}=1$ and $b_{1}=1$ \cite{Gong:2019yxt}. The galaxy bias is treated in a manner consistent with the corresponding treatment of the GW bias in the A and AP models. In the A model, we introduce an independent galaxy bias parameter in each tomographic bin, whereas in the AP model, we adopt a parametrized form for the galaxy bias. We further test the impact of the galaxy bias parametrization and the choice of fiducial galaxy-bias parameters on the constraint on $f^{\mathrm{P}}$ in Sec.~\ref{ssec:GW effective bias}.

\subsection{Cross-correlation angular power spectrum}\label{ssec:angular power spectrum}

In this Section, we follow Ref.~\cite{Pedrotti:2025tfg} and compute the angular power spectra from the perturbative expression for tracer number-density fluctuations,
\begin{equation}\label{eq:9}
C_\ell^{XY}(x_i,x_j)=\frac{2}{\pi}\int\mathrm{d}k\,k^2P(k)\Delta_\ell^{X,x_i}(k)\Delta_\ell^{Y,x_j}(k),
\end{equation}
where $X$ and $Y$ denote different tracers, such as galaxies or GWs, $P(k)$ is the primordial power spectrum, and $\Delta_\ell^{X,x_i}(k)$ is the effective Fourier-space transfer function for tracer $X$ in the $i$th bin, including density, velocity, lensing, and gravitational contributions. Under the Limber approximation, the galaxy-GW cross-correlation, galaxy autocorrelation, and GW autocorrelation spectra can be simplified. The full expressions are given in Appendix A of Ref.~\cite{Pedrotti:2025tfg}. Here we show only the density-density terms, which are directly related to the clustering biases:
\begin{equation}\label{eq:10}
\begin{aligned}
C_{{\ell},\mathrm{den}}^{\mathrm{gGW}}(z_{i},d_{\mathrm{L},j})
=& \int_{0}^{\infty}\mathrm{d}z\,w^{\mathrm{g}}(z,z_{i})\,w^{\mathrm{GW}}[d_{\rm{L}}(z,\lambda),d_{\mathrm{L},j}]\\
&\times\frac{\mathrm{d}d_{\mathrm{L}}}{\mathrm{d}z}(z,\lambda) \frac{H(z)}{cr(z)^{2}}b_{\mathrm{GW}}(z)b_{\mathrm{g}}(z)\\
&\times\mathcal{P}\left[\frac{\ell+1/2}{r(z,\lambda)},z\right],
\end{aligned}
\end{equation}
\begin{equation}\label{eq:11}
\begin{aligned}
C_{{\ell},\mathrm{den}}^{\mathrm{gg}} (z_{i},z_{j})
=& \int_{0}^{\infty}\mathrm{d}z\,w^{\mathrm{g}}(z,z_{i})\,w^{\mathrm{g}}(z,z_{j})\frac{H(z)}{cr(z)^{2}}\\
&\times\left[b_{\mathrm{g}}(z)\right]^2\,\mathcal{P}\left[\frac{\ell+1/2}{r(z,\lambda)},z\right],
\end{aligned}
\end{equation}
\begin{equation}\label{eq:12}
\begin{aligned}
C_{{\ell},\mathrm{den}}^{\mathrm{GWGW}}(d_{\mathrm{L},i},d_{\mathrm{L},j})
=& \int_{0}^{\infty}\mathrm{d}z\,w^{\mathrm{GW}}[d_{\rm{L}}(z,\lambda),d_{\mathrm{L},i}]\\
&\times{w^{\mathrm{GW}}[d_{\rm{L}}(z,\lambda),d_{\mathrm{L},j}]}\\
&\times\left[\frac{\mathrm{d}d_{\mathrm{L}}}{\mathrm{d}z}(z,\lambda)\right]^2\frac{H(z)}{cr(z)^{2}}\\
&\times\left[b_{\mathrm{GW}}(z)\right]^2\,\mathcal{P}\left[\frac{\ell+1/2}{r(z,\lambda)},z\right].
\end{aligned}
\end{equation}
We compute $b_{\mathrm{GW}}(z)$ using Eq.~\eqref{eq:6} for a pure ABH population and Eq.~\eqref{eq:7} for the effective GW clustering bias when ABH and PBH mergers both contribute. For the magnification bias in the lensing terms of both GWs and galaxies, we adopt the parameter choices and numerical results from Ref.~\cite{Du:2025odq}. We use the public code \textsc{pyccl}\footnote{\url{https://ccl.readthedocs.io/en/latest/}} \cite{LSSTDarkEnergyScience:2018yem} with the Limber approximation to compute the theoretical angular power spectra.

For the GW-GW autocorrelation and galaxy-GW cross-correlation, we account for the signal attenuation due to sky localization errors of GW sources, as follows:
\begin{equation}\label{eq:13}
C_\ell^{\mathrm{gGW}}(d_{\mathrm{L},i},z_j)\mapsto C_\ell^{\mathrm{gGW}}(d_{\mathrm{L},i},z_j)\times e^{-\ell(\ell+1)/\ell_{\mathrm{damp}}^2},
\end{equation}
\begin{equation}\label{eq:14}
\begin{aligned}
    C_\ell^{\mathrm{GWGW}}(d_{\mathrm{L},i},d_{\mathrm{L},j})\mapsto & C_\ell^{\mathrm{GWGW}}(d_{\mathrm{L},i},d_{\mathrm{L},j})\\&\times e^{-2\ell(\ell+1)/\ell_{\mathrm{damp}}^2},
\end{aligned}
\end{equation}
where $\ell_{\mathrm{damp}}^2=(2\pi)^{3/2}/\Delta\Omega_{1\sigma}$ and $\Delta\Omega_{1\sigma}$ is the median $1\sigma$ sky-localization uncertainty in each luminosity-distance bin. For the galaxy autocorrelation, $\ell_{\mathrm{max}}$ is limited only by the nonlinear-scale cutoff. For the GW autocorrelation and galaxy-GW cross-correlation, $\ell_{\mathrm{max}}$ is the smaller of the nonlinear-scale cutoff and the angular-resolution limit set by GW localization. The choices of $\ell_{\mathrm{min}}$ and the calculation of $\ell_{\mathrm{max}}$ are detailed in Ref.~\cite{Du:2025odq}. Table~\ref{tab:ell} summarizes the values of $\ell_{\mathrm{min}}$, $\ell_{\mathrm{max}}$, and $\ell_{\mathrm{damp}}$ used in each bin, where the GW luminosity-distance bins correspond to the listed redshift bins.

\begin{figure*}
    \centering
    \includegraphics[width=0.47\linewidth]{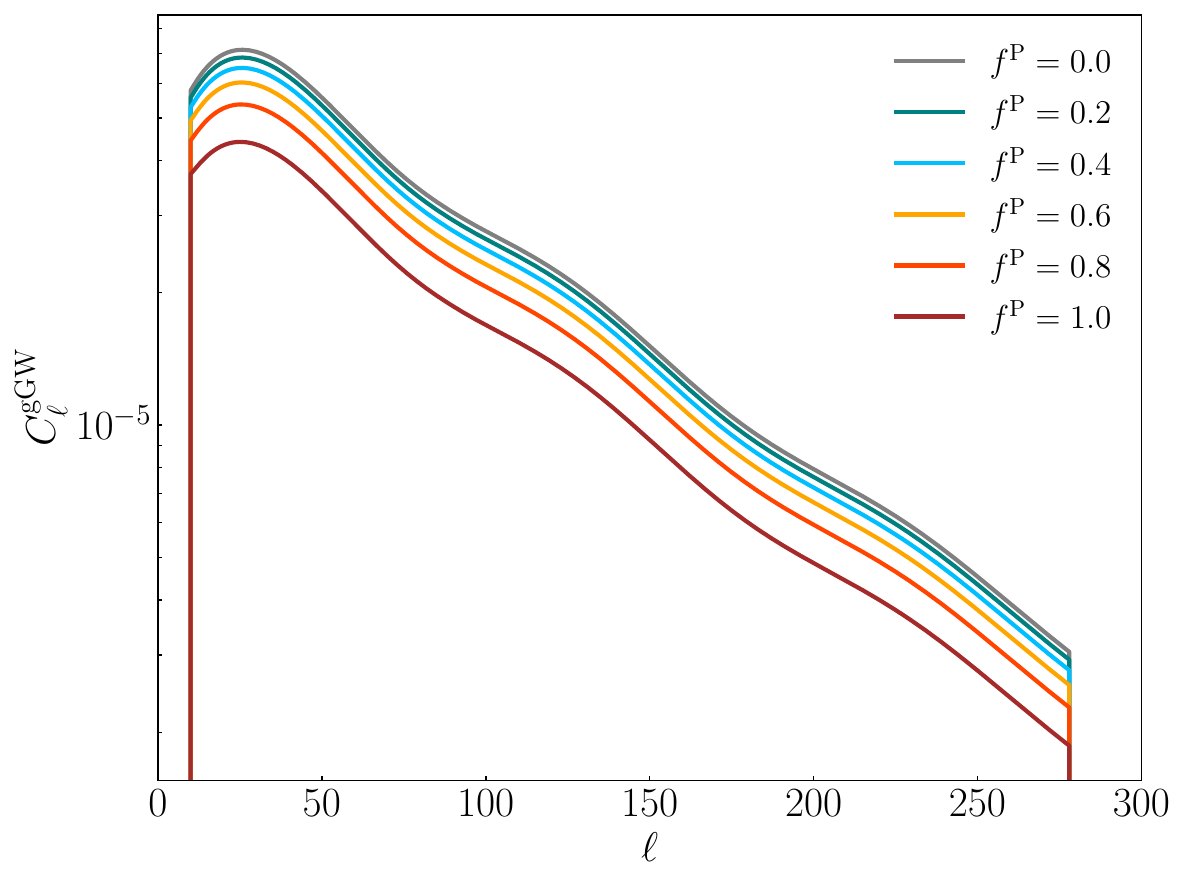}
    \hspace{0.5cm}
    \includegraphics[width=0.47\linewidth]{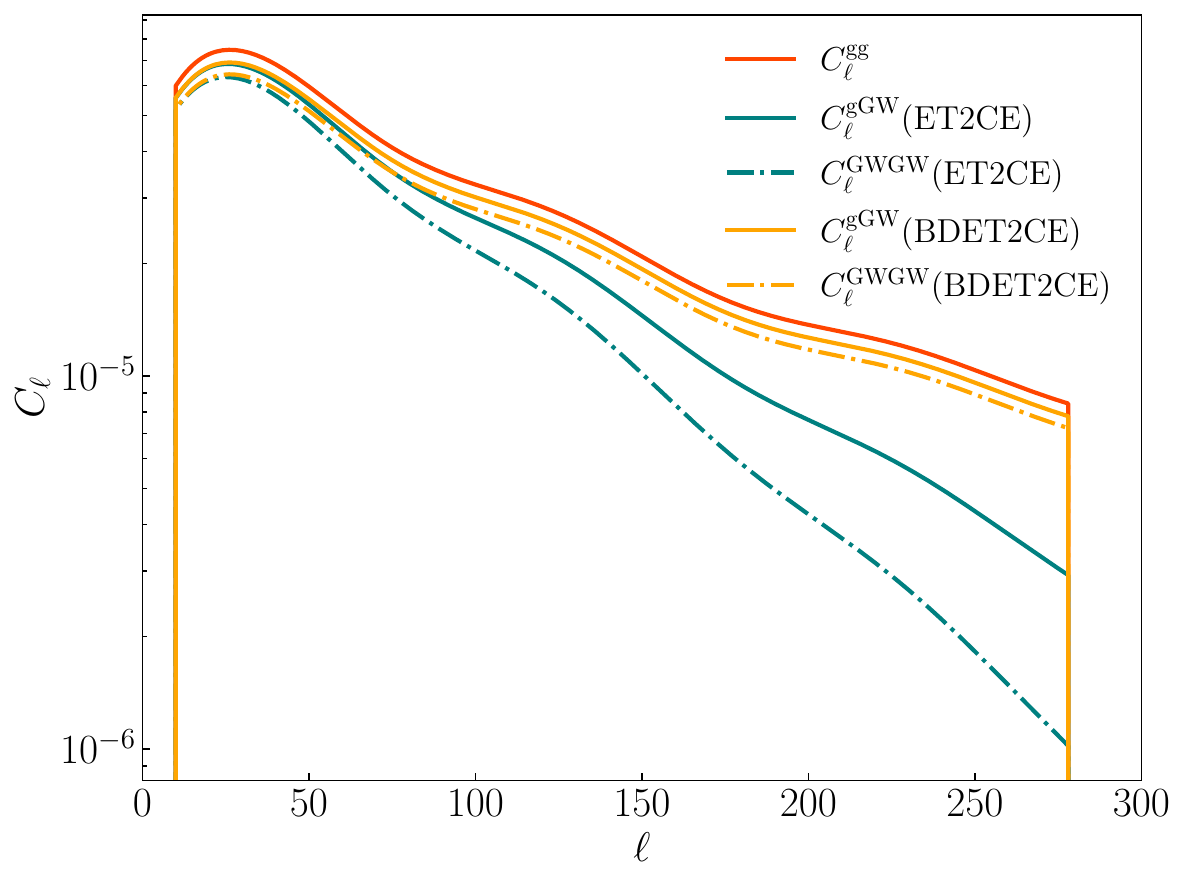}
    \caption{Theoretical angular power spectra with multipole $\ell$ in the seventh redshift bin, $z\in[0.65,0.73]$. The left panel shows the ET2CE galaxy-GW cross-spectrum $C_\ell^{\mathrm{gGW}}$ for different values of $f^{\mathrm{P}}$, illustrating the suppression of the signal as the effective GW bias decreases. The right panel fixes $f^{\mathrm{P}}=0.2$ and compares the galaxy autospectrum $C_\ell^{\mathrm{gg}}$, galaxy-GW cross-spectrum $C_\ell^{\mathrm{gGW}}$, and GW autospectrum $C_\ell^{\mathrm{GWGW}}$ for ET2CE and BDET2CE.}
    \label{fig:Cl_zbin7}
\end{figure*}

Figure~\ref{fig:Cl_zbin7} shows the theoretical angular power spectra as functions of multipole $\ell$ in the seventh redshift bin, $z\in[0.65,0.73]$. The left panel shows the ET2CE galaxy-GW cross-spectrum $C_\ell^{\mathrm{gGW}}$ for different values of $f^{\mathrm{P}}$. Since Eq.~\eqref{eq:10} gives $C_\ell^{\mathrm{gGW}} \propto b_{\mathrm{GW}}$, a larger PBH fraction lowers the effective GW clustering bias and therefore suppresses $C_\ell^{\mathrm{gGW}}$. The right panel fixes $f^{\mathrm{P}}=0.2$ and compares $C_\ell^{\mathrm{gg}}$, $C_\ell^{\mathrm{gGW}}$, and $C_\ell^{\mathrm{GWGW}}$ for ET2CE and BDET2CE. The main advantage of BDET2CE is its sky-localization accuracy, whose error is reduced by two to three orders of magnitude relative to ET2CE \cite{Dong:2024bvw}. This improvement increases $\ell_{\mathrm{damp}}$ and reduces localization damping in the relevant angular power spectra. Because $\ell_{\mathrm{max}}<300$ in this bin whereas $\ell_{\rm damp}\simeq6.5\times10^3\gg\ell_{\rm max}$ for BDET2CE, the damping of $C_\ell^{\mathrm{gGW}}$ and $C_\ell^{\mathrm{GWGW}}$ is almost identical for the BDET2CE network.

The noise is defined as ${\bar{n}_{X,i}}^{-1}$, representing the reciprocal of the total number of observed sources per steradian in the $i$th bin (a redshift bin for galaxies or a luminosity-distance bin for GW sources), given by,
\begin{equation}\label{eq:15}
\mathcal{N}^X(x_i)={\bar{n}_{X,i}}^{-1}=\left[\int_0^\infty\mathrm{d}xW^X(x,x_i)\frac{\mathrm{d}^2N_{\mathrm{obs}}^X}{\mathrm{d}x\mathrm{d}\Omega}\right]^{-1},
\end{equation}
\begin{equation}\label{eq:16}
\begin{aligned}
W^X(x,x_i)=\frac{1}{2}\left\{\operatorname{erf}[u(x^{i+1},x)]-\operatorname{erf}[u(x^i,x)]\right\},
\end{aligned}
\end{equation}
\begin{equation}\label{eq:17}
u(y,x)=\frac{\ln y-\ln x}{\sqrt{2}\sigma_{\ln x}},
\end{equation}
where $\operatorname{erf}$ is the error function, $\sigma_{\ln x}$ is the relative error in the redshift or luminosity-distance measurement, and $x^i$ and $x^{i+1}$ are the lower and upper boundaries of the $i$th bin, respectively. Assuming that the signal and noise are uncorrelated, the observed angular power spectrum is
\begin{equation}\label{eq:18}
\tilde{C}_\ell^{XY}(x_i,x_j)\equiv C_\ell^{XY}(x_i,x_j)+\delta_{XY}\delta_{ij}\mathcal{N}^X(x_i).
\end{equation}

\begin{table*}
\centering
\caption{Multipole ranges and localization damping scales used in the angular power spectrum analysis. For each tomographic bin, defined by the galaxy redshift-bin edges and the corresponding GW luminosity-distance bins, the table lists the adopted $\ell_{\min}$, the nonlinear-scale cutoff $\ell_{\max}^{\mathrm{NL}}$, the localization-limited maximum multipole $\ell_{\max}^{\mathrm{LOC}}$, and the damping scale $\ell_{\mathrm{damp}}$ for ET2CE and BDET2CE.}\label{tab:ell}
\centering
\renewcommand{\arraystretch}{2}
\begin{tabular}{ccccccc}
\hline\hline
\makebox[0.13\textwidth][c]{Bin edges} &
\makebox[0.13\textwidth][c]{$\ell_{\min}$} &
\makebox[0.13\textwidth][c]{$\ell_{\max}^{\mathrm{NL}}$} &
\multicolumn{2}{c}{ET2CE} &
\multicolumn{2}{c}{BDET2CE} \\
\cline{4-5} \cline{6-7}
 &  &  &
\makebox[0.13\textwidth][c]{$\ell_{\max}^{\mathrm{LOC}}$} &
\makebox[0.13\textwidth][c]{$\ell_{\mathrm{damp}}$} &
\makebox[0.13\textwidth][c]{$\ell_{\max}^{\mathrm{LOC}}$} &
\makebox[0.13\textwidth][c]{$\ell_{\mathrm{damp}}$} \\
\hline
0 -- 0.25 & 5 & 40 & 12867 & 1433 & 493018 & 32351 \\
0.25 -- 0.35 & 5 & 103 & 3996 & 944 & 68011 & 20862 \\
0.35 -- 0.43	& 5	& 139 & 3663 & 625 & 59572 & 14010 \\
0.43 -- 0.51	& 5	& 173 & 2526 & 517 & 69093 & 11368 \\
0.51 -- 0.58	& 10 & 207 & 1718 & 418 & 46369 & 9588 \\
0.58 -- 0.65	& 10 & 240 & 1469 & 345 & 28690 & 7753 \\
0.65 -- 0.73	& 10 & 278 & 1142 & 281 & 27331 & 6483 \\
0.73 -- 0.81	& 10 & 319 & 1032 & 255 & 25409 & 6107 \\
0.81 -- 0.90	& 15 & 366 & 995 & 219 & 23328 & 5370 \\
0.90 -- 1.00	& 15 & 421 & 815 & 198 & 18054 & 4676 \\
1.00 -- 1.11	& 15 & 486 & 673 & 170 & 17330 & 4091 \\
1.11 -- 1.25	& 15 & 569 & 613 & 147 & 15624 & 3606 \\
1.25 -- 1.44	& 20 & 587 & 512 & 125 & 12340 & 3124 \\
1.44 -- 1.74	& 20 & 884 & 447 & 105 & 11523 & 2714 \\
1.74 -- 4.00	& 20 & 2386 & 267 & 67 & 7327 & 1944 \\
\hline\hline
\end{tabular}
\end{table*}

\subsection{Signal-to-noise ratios and Fisher analysis}\label{ssec:SNRs and FIM}
We use signal-to-noise ratios (SNRs) and FIM to assess the ability of future GW observations to identify a PBH contribution.

Following Ref.~\cite{Libanore:2023ovr}, we define SNR by comparing the angular power spectra predicted by the AP and A models,
\begin{equation}\label{eq:19}
\mathrm{SNR}^2 = f_{\mathrm{sky}}\sum_\ell(2\ell+1) \left( \tilde{\mathbf{C}}_{\ell} - \tilde{\mathbf{C}}_{\ell}^{\mathrm{AP}} \right)^T \mathcal{M}_{\ell}^{-1} \left( \tilde{\mathbf{C}}_{\ell} - \tilde{\mathbf{C}}_{\ell}^{\mathrm{AP}} \right),
\end{equation}
where $f_{\mathrm{sky}}=0.42$ is the sky coverage fraction of the CSST photometric survey. The vector $\tilde{\mathbf{C}}_{\ell}$ contains the observed angular power spectra in the A model, while $\tilde{\mathbf{C}}_{\ell}^{\mathrm{AP}}$ contains the corresponding spectra in the AP model. We use the shorthand $\tilde{C}_\ell^{XY}(x_i,x_j) = \tilde{C}_{\ell,ij}^{XY}$. The vector $\tilde{\mathbf{C}}_{\ell}$ is constructed as
\begin{equation}\label{eq:20}
\begin{split}
\tilde{\mathbf{C}}_{\ell} = (&\, \tilde{C}_{\ell,11}^{XX}\ \ \tilde{C}_{\ell,12}^{XX}\ \ \ldots \ \tilde{C}_{\ell,NN}^{XX}\ \ \tilde{C}_{\ell,11}^{XY}\ \ \ldots \\
& \tilde{C}_{\ell,NM}^{XY}\ \ \tilde{C}_{\ell,11}^{YY}\ \ \ldots \ \tilde{C}_{\ell,MM}^{YY}),
\end{split}
\end{equation}
where $N=15$ is the number of galaxy redshift bins and $M=15$ is the number of GW luminosity-distance bins. The luminosity-distance bin edges are obtained by converting the galaxy redshift-bin edges using the fiducial Planck 2018 $\Lambda$CDM cosmology. The covariance matrix $\mathcal{M}_{\ell}$ is
\begin{equation}\label{eq:21}
\begin{aligned}
\mathcal{M}_{\ell,IJ}
=& \left( C_{\ell,ip}^{XX} + \frac{\delta_{ip}}{\bar{n}_{X,i}} \right) \left( C_{\ell,jq}^{YY} + \frac{\delta_{jq}}{\bar{n}_{Y,j}} \right)\\
& + \left( C_{\ell,iq}^{XY} + \frac{\delta_{iq}\delta_{XY}}{\bar{n}_{X,i}} \right) \left( C_{\ell,jp}^{XY} + \frac{\delta_{jp}\delta_{XY}}{\bar{n}_{X,j}} \right),
\end{aligned}
\end{equation}
where $IJ$ corresponds to the index combination of $(X,ij)$ and $(Y,pq)$ in the $\tilde{\mathbf{C}}_{\ell}$ vector.

We use the FIM approach to estimate the uncertainties of the bias parameters. For a parameter set $\{\theta_\alpha\}$, the Fisher matrix elements are given by \cite{Pedrotti:2025tfg}
\begin{equation}\label{eq:22}
\begin{aligned}
    F_{\alpha\beta}=f_\mathrm{sky}\sum_\ell\frac{2\ell+1}{2}\mathrm{Tr}[\mathcal{C}_\ell^{-1}(\partial_\alpha\mathcal{C}_\ell)\mathcal{C}_\ell^{-1}(\partial_\beta\mathcal{C}_\ell)],
\end{aligned}
\end{equation}
\begin{equation}\label{eq:23}
\begin{aligned}
    &\mathcal{C}_\ell =\\
&\resizebox{\columnwidth}{!}{$
\begin{pmatrix}
\tilde{C}_\ell^{\rm gg}(z_1,z_1) & \cdots & \tilde{C}_\ell^{\rm gg}(z_1,z_N)
 & \tilde{C}_\ell^{\rm gGW}(z_1,d_{\mathrm{L},1}) & \cdots
 & \tilde{C}_\ell^{\rm gGW}(z_1,d_{\mathrm{L},M}) \\[4pt]
 & \ddots & \vdots & \vdots & & \vdots \\[4pt]
 & & \tilde{C}_\ell^{\rm gg}(z_N,z_N)
 & \tilde{C}_\ell^{\rm gGW}(z_N,d_{\mathrm{L},1}) & \cdots
 & \tilde{C}_\ell^{\rm gGW}(z_N,d_{\mathrm{L},M}) \\[4pt]
 & & & \tilde{C}_\ell^{\rm GWGW}(d_{\mathrm{L},1},d_{\mathrm{L},1})
 & \cdots & \tilde{C}_\ell^{\rm GWGW}(d_{\mathrm{L},1},d_{\mathrm{L},M}) \\[4pt]
 & & & & \ddots & \vdots \\[4pt]
 & & & & & \tilde{C}_\ell^{\rm GWGW}(d_{\mathrm{L},M},d_{\mathrm{L},M})
\end{pmatrix}
$},
\end{aligned}
\end{equation}
where $\mathcal{C}_\ell$ is the covariance matrix containing the auto- and cross-correlation angular power spectra. In the A model bias analysis, the parameter set is $\theta_\alpha = \{b_{\mathrm{g},1}, b_{\mathrm{g},2}, \dots, b_{\mathrm{g},N}, b_{\mathrm{GW},1}, b_{\mathrm{GW},2}, \dots, b_{\mathrm{GW},M}\}$. The fiducial galaxy and GW bias values are evaluated at the center of each bin using Eqs.~\eqref{eq:8} and \eqref{eq:6}, respectively. The derivative $\partial_\alpha\mathcal{C}_\ell$ is computed by varying one parameter at a time while keeping all other parameters fixed,
\begin{equation}\label{eq:24}
\begin{split}
\partial_\alpha\mathcal{C}_\ell=\frac{\partial\mathcal{C}_\ell}{\partial\theta_\alpha} = \frac{\mathcal{C}_\ell(\theta_\alpha+\Delta\theta_\alpha)-\mathcal{C}_\ell(\theta_\alpha-\Delta\theta_\alpha)}{2\Delta\theta_\alpha},
\end{split}
\end{equation}
where $\Delta\theta_\alpha$ is a small perturbation around the fiducial value of $\theta_\alpha$. We set $\Delta\theta_\alpha=f_\alpha\theta_\alpha$, with $f_{b_{\mathrm{g}}}=1.0\times10^{-5}$ for the galaxy bias parameters and $f_{b_{\mathrm{GW}}}=1.0\times10^{-5}$ for the GW bias parameters. The marginalized $1\sigma$ uncertainty of $\theta_\alpha$ is then $\sigma_{\theta_\alpha}=\sqrt{(F^{-1})_{\alpha\alpha}}$.

We then consider the AP model, in which ABH and PBH mergers both contribute, and use the FIM to forecast constraints on the PBH fraction. In this case, we jointly vary five parameters, $\theta_\alpha = \{C, D, b_0, b_1, f^{\mathrm{P}}\}$. The ABH bias parameters in Eq.~\eqref{eq:6} have fiducial values $C=1.2$ and $D=0.59$, while the galaxy bias parameters in Eq.~\eqref{eq:8} have fiducial values $b_0=1$ and $b_1=1$. The parameter $f^{\mathrm{P}}$ denotes the PBH fraction in the local total merger rate, and we consider fiducial values $f^{\mathrm{P}}=0.2, 0.4, 0.6$, and $0.8$. For all five parameters, we set the relative finite-difference step to $f_\alpha=1.0\times10^{-3}$.

\section{Results and discussion}\label{sec:results}

In this Section, we present cross-correlation forecasts for the CSST photometric galaxy sample combined with two GW detector networks: the third-generation ground-based ET2CE network and the multiband BDET2CE network.

We assess the PBH contribution in three complementary ways. We first compute SNR for distinguishing the AP model from the A model, which sets the overall detectability of a PBH contribution (Sec.~\ref{ssec:snr}). We then forecast the GW clustering bias in the A model and identify the PBH fraction at which the AP-model effective bias departs from it, marking where a PBH contribution becomes identifiable (Sec.~\ref{ssec:GW bias}). Finally, we treat the PBH fraction $f^{\mathrm{P}}$ as a free parameter and forecast how precisely it can be measured (Sec.~\ref{ssec:GW effective bias}). The first two analyses quantify whether a PBH contribution can be detected, whereas the third quantifies how well its amplitude can be measured; as shown below, a contribution can be detected at substantially lower fractions than it can be precisely measured.

\subsection{Signal-to-noise ratios}\label{ssec:snr}

\begin{figure*}
    \centering
    \includegraphics[width=1\linewidth]{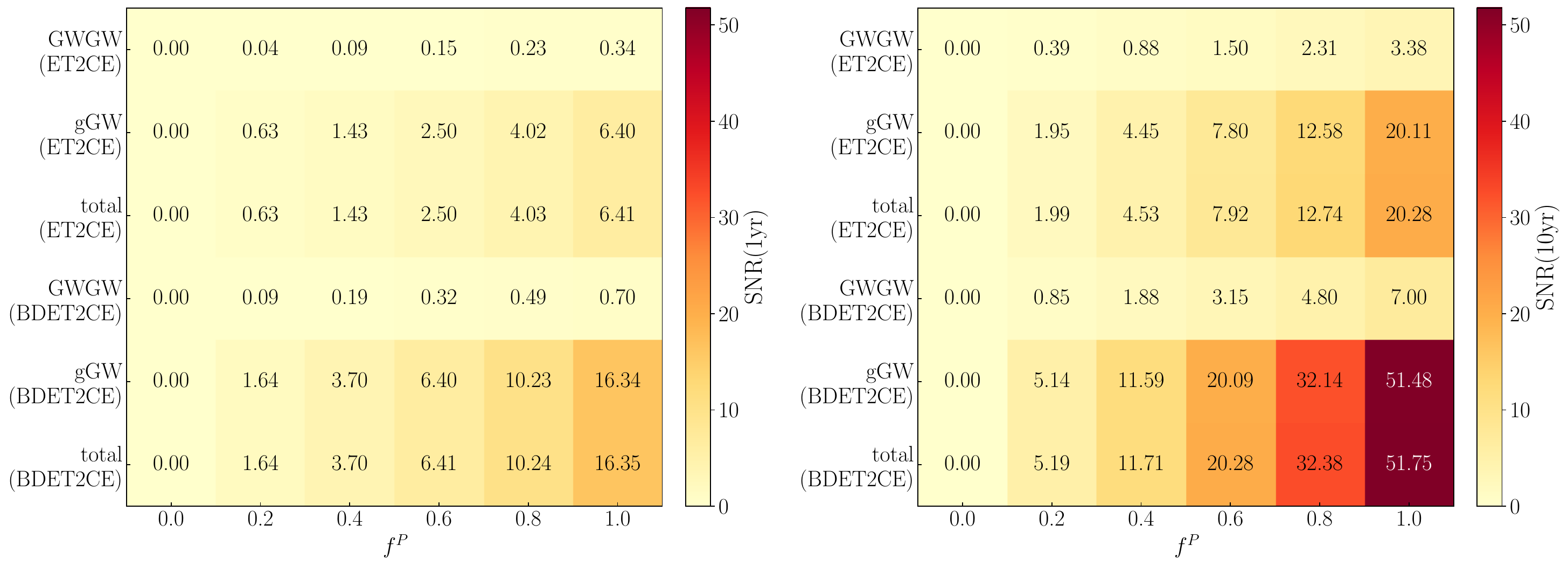}
    \caption{SNR for distinguishing the AP model from the A model as a function of $f^{\mathrm{P}}$. The left and right panels show 1-year and 10-year observations, respectively. Curves correspond to the GW autocorrelation (GWGW), the galaxy-GW cross-correlation (gGW), and the joint three-spectrum analysis including the galaxy autocorrelation (total).}
    \label{fig:SNR_fp}
\end{figure*}

We first compute SNR for the GW autocorrelation alone, the galaxy-GW cross-correlation alone, and the joint analysis that also includes the galaxy autocorrelation, for different values of $f^{\mathrm{P}}$. Figure~\ref{fig:SNR_fp} shows SNR as a function of $f^{\mathrm{P}}$ for ET2CE and BDET2CE.

Both the choice of observable and the detector network strongly affect SNR. The galaxy-GW cross-correlation is far more sensitive than the GW autocorrelation: for 10 years of ET2CE observations at $f^{\mathrm{P}}=0.2$, the cross-correlation gives $\mathrm{SNR}=1.95$, against only 0.39 for the GW autocorrelation. This gap arises because the cross-correlation pairs the sparse GW sample with the dense, precisely located CSST galaxies, so it is far less affected by GW shot noise and, from Eqs.~\eqref{eq:13} and \eqref{eq:14}, the localization damping enters its spectrum once rather than twice. Extending the observation from 1 to 10 years raises SNR by about $\sqrt{10}$, reflecting the tenfold growth in the number of detected events and the corresponding drop in GW shot noise; for the ET2CE cross-correlation at $f^{\mathrm{P}}=1.0$, for example, SNR grows from 6.40 to 20.11.

The largest gain comes from the multiband network. B-DECIGO shrinks the sky-localization area of GW sources by two to three orders of magnitude, which raises the damping multipole $\ell_{\mathrm{damp}}$ by more than an order of magnitude (Table~\ref{tab:ell}). Because localization errors suppress the GW spectra above $\ell_{\mathrm{damp}}$, the ET2CE cross- and autospectra are localization limited over much of the redshift range, whereas for BDET2CE $\ell_{\mathrm{damp}}$ lies close to or above the nonlinear-scale cutoff and the spectra stay essentially undamped over the usable range. The small-scale, high-$\ell$ modes that BDET2CE thereby recovers add to the SNR sum and raise the cross-correlation SNR by a nearly constant factor of about 2.6, from 1.95 to 5.14 at $f^{\mathrm{P}}=0.2$ and from 20.11 to 51.48 at $f^{\mathrm{P}}=1.0$. As a result, with 10 years of data the CSST cross-correlation reaches the $\mathrm{SNR}>4$ detection level at $f^{\mathrm{P}}\simeq0.4$ for ET2CE and already at $f^{\mathrm{P}}\simeq0.2$ for BDET2CE, while the GW autocorrelation alone stays below 4 for ET2CE at all fractions and needs $f^{\mathrm{P}}\gtrsim0.8$ even for BDET2CE. Multiband localization therefore substantially lowers the PBH fraction at which a contribution becomes detectable.

We also evaluate the total SNR from all three spectra: the GW autocorrelation, the galaxy-GW cross-correlation, and the galaxy autocorrelation. SNR in Eq.~\eqref{eq:19} is built from the difference between the A and AP models. Because the galaxy autocorrelation does not depend on the PBH fraction, it is identical in the two models and does not help distinguish them; the galaxy-GW cross-correlation therefore carries essentially all of the discriminating power, and the total SNR is only marginally higher than the cross-correlation alone. We nonetheless retain the galaxy autocorrelation because, in the Fisher forecasts below, it tightly constrains the galaxy bias and breaks its degeneracy with the GW bias.

\begin{table*}
\centering
\caption{Relative $1\sigma$ uncertainties of the galaxy and GW clustering bias parameters in the A model. The constraints are obtained from the joint analysis of the galaxy autospectrum, GW autospectrum, and galaxy-GW cross-spectrum for different GW detector configurations and observing times. All entries are percentages.}\label{tab:results}
\centering
\renewcommand{\arraystretch}{2}
\begin{tabular}{cccccccccccccccc}
\hline\hline
\makebox[0.13\textwidth][c]{Galaxy bias} &
\makebox[0.13\textwidth][c]{CSST} &
\makebox[0.13\textwidth][c]{GW bias} &
\makebox[0.13\textwidth][c]{ET2CE(1yr)} &
\makebox[0.13\textwidth][c]{ET2CE(10yr)} &
\makebox[0.13\textwidth][c]{BDET2CE(1yr)} &
\makebox[0.13\textwidth][c]{BDET2CE(10yr)} \\
\hline
$\Delta b_{\mathrm{g},1}/b_{\mathrm{g},1}$ & 2.68\% & $\Delta b_{\mathrm{GW},1}/b_{\mathrm{GW},1}$ & 83.84\% & 26.65\% & 83.72\% & 26.62\% \\
$\Delta b_{\mathrm{g},2}/b_{\mathrm{g},2}$ & 1.07\% & $\Delta b_{\mathrm{GW},2}/b_{\mathrm{GW},2}$ & 41.73\% & 13.23\% & 41.51\% & 13.17\% \\
$\Delta b_{\mathrm{g},3}/b_{\mathrm{g},3}$ & 0.80\%	& $\Delta b_{\mathrm{GW},3}/b_{\mathrm{GW},3}$ & 31.70\% & 10.05\% & 31.13\% & 9.87\% \\
$\Delta b_{\mathrm{g},4}/b_{\mathrm{g},4}$ & 0.65\%	& $\Delta b_{\mathrm{GW},4}/b_{\mathrm{GW},4}$ & 26.02\% & 8.25\% & 25.01\% & 7.93\% \\
$\Delta b_{\mathrm{g},5}/b_{\mathrm{g},5}$ & 0.54\% & $\Delta b_{\mathrm{GW},5}/b_{\mathrm{GW},5}$ & 22.69\% & 7.20\% & 20.91\% & 6.63\% \\
$\Delta b_{\mathrm{g},6}/b_{\mathrm{g},6}$ & 0.47\% & $\Delta b_{\mathrm{GW},6}/b_{\mathrm{GW},6}$ & 20.95\% & 6.64\% & 18.04\% & 5.72\% \\
$\Delta b_{\mathrm{g},7}/b_{\mathrm{g},7}$ & 0.41\% & $\Delta b_{\mathrm{GW},7}/b_{\mathrm{GW},7}$ & 20.46\% & 6.48\% & 15.73\% & 4.99\% \\
$\Delta b_{\mathrm{g},8}/b_{\mathrm{g},8}$ & 0.37\% & $\Delta b_{\mathrm{GW},8}/b_{\mathrm{GW},8}$ & 19.89\% & 6.30\% & 13.84\% & 4.39\% \\
$\Delta b_{\mathrm{g},9}/b_{\mathrm{g},9}$ & 0.33\% & $\Delta b_{\mathrm{GW},9}/b_{\mathrm{GW},9}$ & 20.55\% & 6.51\% & 12.34\% & 3.91\% \\
$\Delta b_{\mathrm{g},10}/b_{\mathrm{g},10}$ & 0.30\% & $\Delta b_{\mathrm{GW},10}/b_{\mathrm{GW},10}$ & 20.87\% & 6.61\% & 11.04\% & 3.50\% \\
$\Delta b_{\mathrm{g},11}/b_{\mathrm{g},11}$ & 0.27\% & $\Delta b_{\mathrm{GW},11}/b_{\mathrm{GW},11}$ & 22.29\% & 7.06\% & 9.96\% & 3.16\% \\
$\Delta b_{\mathrm{g},12}/b_{\mathrm{g},12}$ & 0.24\% & $\Delta b_{\mathrm{GW},12}/b_{\mathrm{GW},12}$ & 24.28\% & 7.70\% & 9.16\% & 2.91\% \\
$\Delta b_{\mathrm{g},13}/b_{\mathrm{g},13}$ & 0.23\% & $\Delta b_{\mathrm{GW},13}/b_{\mathrm{GW},13}$ & 27.97\% & 8.87\% & 8.62\% & 2.74\% \\
$\Delta b_{\mathrm{g},14}/b_{\mathrm{g},14}$ & 0.23\% & $\Delta b_{\mathrm{GW},14}/b_{\mathrm{GW},14}$ & 34.61\% & 10.99\% & 8.59\% & 2.73\% \\
$\Delta b_{\mathrm{g},15}/b_{\mathrm{g},15}$ & 0.23\% & $\Delta b_{\mathrm{GW},15}/b_{\mathrm{GW},15}$ & 78.50\% & 25.11\% & 11.06\% & 3.53\% \\
\hline\hline
\end{tabular}
\end{table*}

\subsection{Constraints on GW bias in the A model}\label{ssec:GW bias}

We now present the joint constraints on the galaxy and GW clustering bias parameters in the A model, obtained from the combined analysis of the GW autocorrelation, galaxy autocorrelation, and galaxy-GW cross-correlation spectra. The results are summarized in Table~\ref{tab:results}.

The CSST photometric survey constrains the galaxy clustering bias extremely well, with relative errors $\Delta b_{\mathrm{g},i}/b_{\mathrm{g},i}$ below 3\% in all 15 redshift bins. The GW clustering bias is much harder to measure, and its precision depends strongly on the detector network and the observing time. With redshift, the GW bias precision first improves and then degrades, because two effects compete: the number of detected GW events grows with redshift, while their sky localization worsens. For ET2CE with one year of observation, the GW bias errors are large, from 83.84\% in the first bin to 19.89\% in the best-constrained bin $b_{\mathrm{GW},8}$ at $z\sim0.73$--$0.81$, where the two effects balance. Extending ET2CE to 10 years adds ten times as many events and reduces the errors by about 68\% (the $\sqrt{10}$ scaling of shot noise), bringing them below 7\% for $z\sim0.58$--$1.00$. The multiband BDET2CE network localizes high-redshift events far better, removing the localization penalty at high redshift; the best-constrained bin therefore moves up to $b_{\mathrm{GW},14}$ at $z\sim1.44$--$1.74$, where the error falls to 2.73\%, about 75.16\% smaller than ET2CE in that bin. BDET2CE keeps the GW bias error below 7\% across $z\sim0.51$--$4.00$, and since this range is set by our $z\sim4$ analysis cutoff rather than by a loss of sensitivity, a deeper survey would likely extend it. These results show that the sky localization of GW sources is the key factor controlling how well their clustering bias can be measured.

\begin{figure*}
    \centering
    \includegraphics[width=1\linewidth]{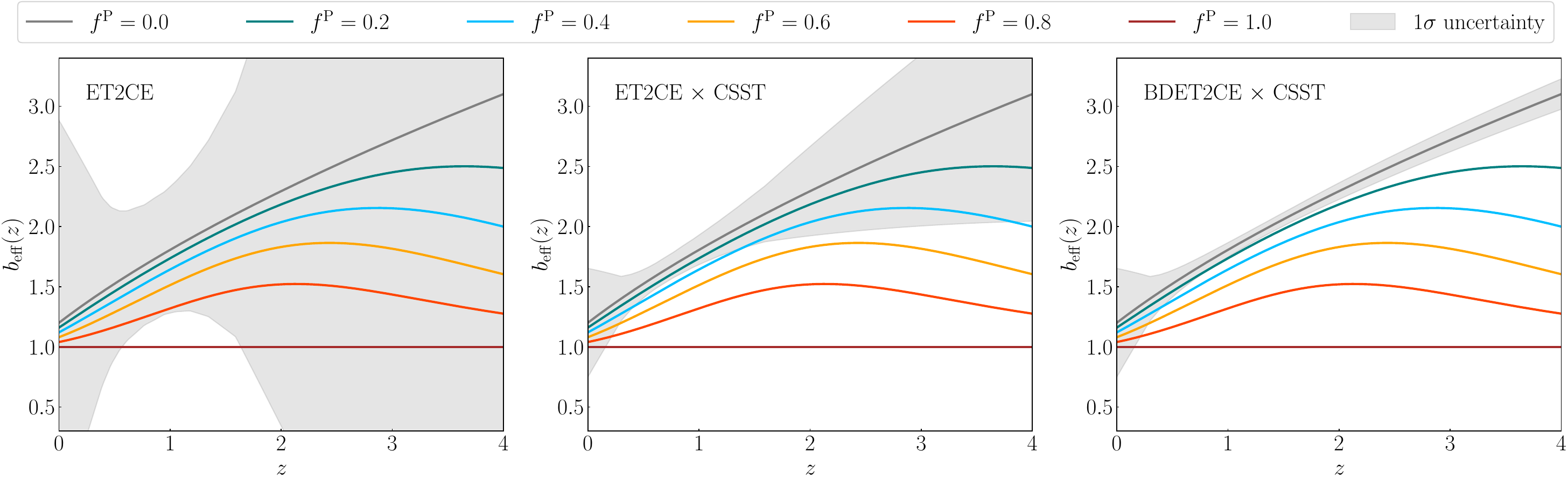}
    \caption{GW effective bias compared with the $1\sigma$ A model bias constraints for 10-year observations. The panels show ET2CE GW autocorrelation only (left), ET2CE-CSST joint analysis (middle), and BDET2CE-CSST joint analysis (right). The gray bands show the A model $1\sigma$ uncertainties, while the colored curves show AP model effective biases for different PBH fractions. We note that the effective-bias curves shown here assume $b^{\mathrm{P}}=1$. If $b^{\mathrm{P}}$ is allowed to vary, the colored curves will shift accordingly.}
    \label{fig:b_eff_bins_errbar}
\end{figure*}

Figure~\ref{fig:b_eff_bins_errbar} turns these bias measurements into a detection statement. In each panel, the gray band is the $1\sigma$ uncertainty on the GW bias measured under the A model, while the colored curves show the effective bias $b_{\mathrm{eff}}(z)$ predicted by the AP model for different $f^{\mathrm{P}}$. Because PBHs carry a low, nearly constant bias, a larger $f^{\mathrm{P}}$ pulls $b_{\mathrm{eff}}$ below the pure-ABH ($f^{\mathrm{P}}=0$) curve, and the curves fan out increasingly toward high redshift, where PBH mergers make up a growing share of the total rate. Once a curve leaves the band, we regard the PBH contribution as becoming distinguishable from the pure-ABH expectation. The detectable $f^{\mathrm{P}}$ is therefore set by the width of the band, that is, by how well the GW bias is measured. With the ET2CE GW autocorrelation alone (left panel), the band is so wide that even high PBH fractions remain consistent with it, so GW data alone are sensitive only when PBH mergers dominate. Including CSST in the joint analysis (middle panel) shrinks the band, and the curves leave it for $f^{\mathrm{P}}\gtrsim0.4$. BDET2CE shrinks it further and keeps it narrow out to high redshift, where the ET2CE band instead broadens and where the curves fan out most; this combination pushes the detectable fraction down to $f^{\mathrm{P}}\simeq0.2$ (right panel).

Our baseline analysis assumes $b^{\mathrm{P}}=1$. However, the choice of $b^{\mathrm{P}}$ affects the detection threshold of $f^{\mathrm{P}}$. Within the range considered here, increasing $b^{\mathrm{P}}$ reduces the bias difference between the PBH and ABH populations, particularly at high redshift, thereby shifting the effective-bias curves of the AP model toward the gray uncertainty band of the A model and raising the PBH detection threshold; conversely, decreasing $b^{\mathrm{P}}$ lowers the threshold. We further test this dependence using the joint analysis of ET2CE and CSST. The detection threshold is about $f^{\mathrm{P}}\simeq0.3$ for $b^{\mathrm{P}}=0.5$, $f^{\mathrm{P}}\simeq0.4$ for the fiducial $b^{\mathrm{P}}=1$, and $f^{\mathrm{P}}\simeq0.5$ for $b^{\mathrm{P}}=1.4$. Therefore, if future simulations or observations indicate that the bias of early PBHs deviates significantly from unity, the corresponding detection threshold would need to be revised accordingly.

\subsection{Constraints on the PBH fraction in the AP model}\label{ssec:GW effective bias}

We next use the FIM to jointly constrain the parameters entering the GW effective bias, $C$, $D$, and $f^{\mathrm{P}}$, together with the galaxy bias parameters $b_0$ and $b_1$, in the AP model. The main goal is to quantify how well future joint observations can constrain the PBH fraction $f^{\mathrm{P}}$.

Table~\ref{tab:f_P} gives the $1\sigma$ relative uncertainty on $f^{\mathrm{P}}$ for different GW detector configurations and observing times, using the joint analysis of the three angular power spectra. ET2CE does not constrain $f^{\mathrm{P}}$ with one year of observation and provides only a weak constraint, 88.63\%, for the high-fraction case $f^{\mathrm{P}}=0.8$ after ten years. BDET2CE performs better: it reaches 78.21\% for $f^{\mathrm{P}}=0.8$ after one year, and after ten years it reaches 91.48\%, 43.16\%, 32.30\%, and 24.88\% for $f^{\mathrm{P}}=0.2$, 0.4, 0.6, and 0.8, respectively. These results show that the multi-band BDET2CE network improves the direct constraint on $f^{\mathrm{P}}$ and lowers the PBH fraction at which useful constraints become possible. These direct constraints on $f^{\mathrm{P}}$ are weaker than the detection thresholds found above, and the two should not be conflated. Obtaining evidence for a nonzero PBH contribution, quantified by SNR (Sec.~\ref{ssec:snr}) and by the departure of the GW bias from the pure-ABH expectation (Sec.~\ref{ssec:GW bias}), is statistically easier than measuring $f^{\mathrm{P}}$ as a free parameter. For BDET2CE, a PBH contribution is already detectable near $f^{\mathrm{P}}=0.2$, whereas $f^{\mathrm{P}}$ itself is constrained only at the 91.48\% level there, improving to 43.16\% at $f^{\mathrm{P}}=0.4$.

\begin{figure}
    \centering
    \includegraphics[width=1\linewidth]{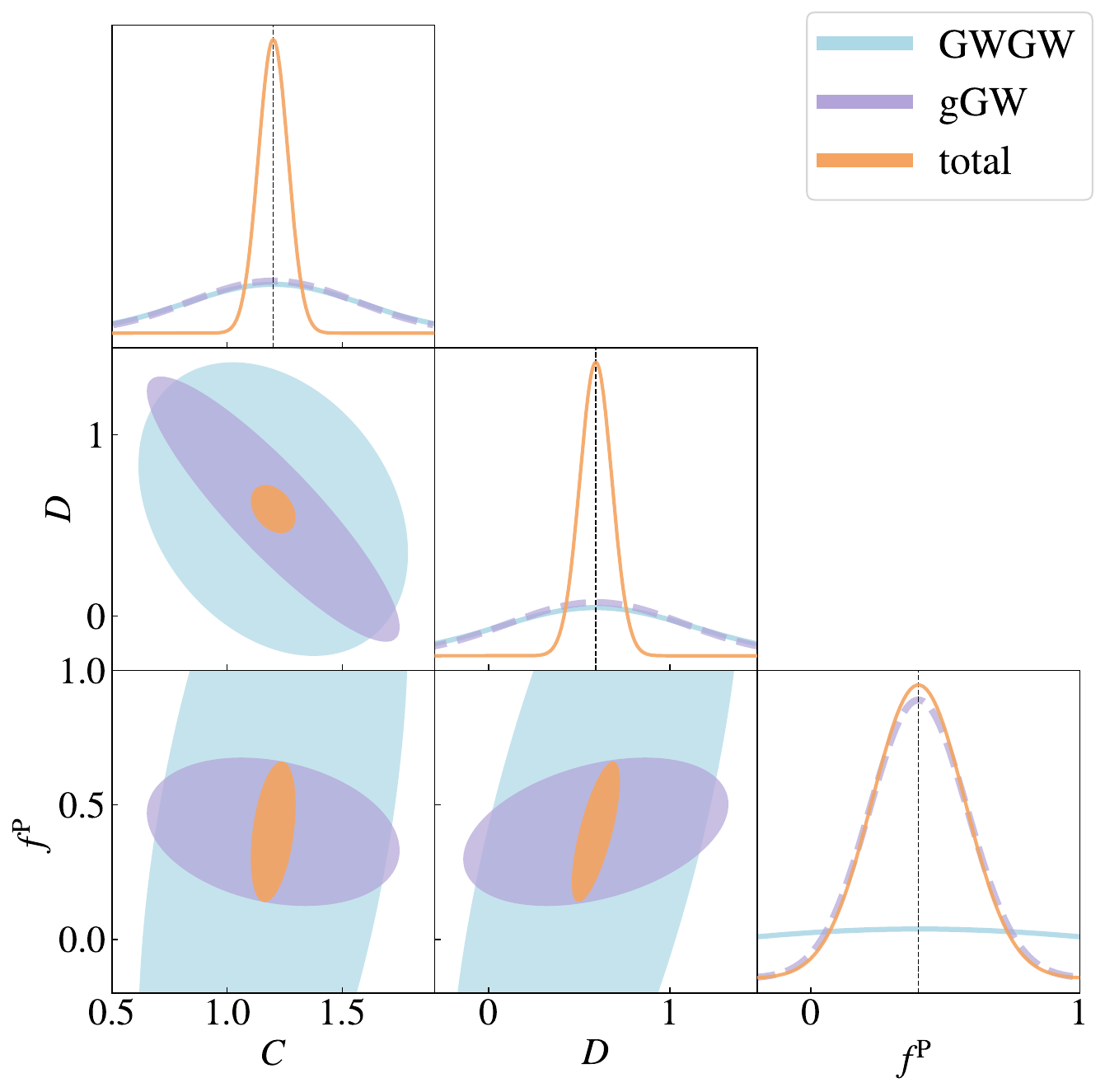}
    \caption{$1\sigma$ posterior constraints on the AP model parameters $C$, $D$, and $f^{\mathrm{P}}$ for the fiducial case $f^{\mathrm{P}}=0.4$. The constraints use the 10-year BDET2CE GW sample and the CSST photometric galaxy sample. Results are compared for the GW autocorrelation only (GWGW), the galaxy-GW cross-correlation only (gGW), and the joint analysis including the galaxy autocorrelation (total).}
    \label{fig:fp4_BDET2CE_10yr_corner}
\end{figure}

We further dissect where the constraining power comes from, taking the AP model with $f^{\mathrm{P}}=0.4$ and 10 years of BDET2CE observations combined with CSST (Table~\ref{tab:f_P_0.4}; Fig.~\ref{fig:fp4_BDET2CE_10yr_corner} shows the posteriors of $C$, $D$, and $f^{\mathrm{P}}$). The GW autocorrelation alone barely constrains $f^{\mathrm{P}}$, which remains positively correlated with the ABH bias parameters $C$ and $D$. Adding the galaxy-GW cross-correlation tightens $f^{\mathrm{P}}$ to 45.48\% and slightly improves $C$ and $D$, because the cross-correlation links the GW bias to the precisely mapped galaxy field. The decisive step is including the galaxy autocorrelation, which collapses the errors on the galaxy bias parameters $b_0$ and $b_1$ from 26.60\% and 41.37\% to 0.27\% and 0.32\%. It can do so because it measures the galaxy clustering amplitude directly through $C_\ell^{\mathrm{gg}}\propto b_{\mathrm{g}}^2$, using the dense CSST sample over all bins and all multipoles up to the nonlinear cutoff, free from the GW shot noise and localization damping that limit the GW spectra. Since the galaxy and GW biases are degenerate in the cross-correlation, pinning down $b_0$ and $b_1$ in turn tightens $C$ and $D$ to 5.35\% and 14.87\%. The PBH fraction $f^{\mathrm{P}}$ itself improves only modestly, as it is nearly independent of the galaxy bias, but the galaxy autocorrelation still strengthens the parameter set as a whole. This shows why the three-spectrum combination is essential: the spectra carry complementary information, and together they break the degeneracies that limit any one of them alone.

To test the robustness of the constraint on $f^{\mathrm{P}}$ to the modeling of galaxy bias, we perform two additional checks. First, following the same treatment as in the A model, we perform a Fisher analysis with 18 parameters, $\theta_\alpha = \{C, D, f^{\mathrm{P}}, b_{\mathrm{g},1}, b_{\mathrm{g},2}, \dots, b_{\mathrm{g},15}\}$, to reassess the constraint on $f^{\mathrm{P}}$. Even with this more conservative scheme, the relative $1\sigma$ uncertainty on $f^{\mathrm{P}}$ remains almost unchanged compared with the results in Table~\ref{tab:f_P}, increasing by only about 0.04 to 0.07 percentage points. This is because the CSST galaxy autocorrelation angular power spectrum provides strong constraints on the galaxy bias parameters whether they are parametrized by Eq.~\eqref{eq:8} or left free in each bin. Second, we further test the dependence of the results on the fiducial galaxy-bias parameters. For this test, we adopt the 10-year BDET2CE observations combined with CSST and set $f^{\mathrm{P}}=0.4$ as the fiducial PBH fraction. Keeping the functional form $b_{\mathrm{g}}(z) = b_{0}(1+z)^{b_{1}}$ unchanged, we consider four cases, $(b_0,b_1)=(1.2,1.0)$, $(0.8,1.0)$, $(1.0,1.2)$, and $(1.0,0.8)$. The corresponding relative $1\sigma$ uncertainties on $f^{\mathrm{P}}$ are 41.23\%, 46.17\%, 40.74\%, and 46.41\%, respectively, compared with 43.16\% for the fiducial case $(b_0,b_1)=(1.0,1.0)$. The maximum absolute change relative to the fiducial result is about 3.25 percentage points. Thus, within the range considered here, our forecasted constraint on $f^{\mathrm{P}}$ is reasonably robust to both the galaxy-bias parametrization and the choice of fiducial bias parameters.

\begin{table}
\centering
\caption{Relative $1\sigma$ uncertainty $\Delta f^{\mathrm{P}}/f^{\mathrm{P}}$ for different PBH fractions in the AP model under different GW detector configurations and observation times, based on the joint three-spectrum analysis, including $C_\ell^{\mathrm{gg}}$, $C_\ell^{\mathrm{gGW}}$, and $C_\ell^{\mathrm{GWGW}}$.}\label{tab:f_P}
\centering
\centering
\renewcommand{\arraystretch}{2}
\begin{tabular}{cccccc}
\hline\hline
\makebox[0.072\textwidth][c]{$f^{\mathrm{P}}$} & \makebox[0.072\textwidth][c]{0.2} & \makebox[0.072\textwidth][c]{0.4} & \makebox[0.072\textwidth][c]{0.6} & \makebox[0.072\textwidth][c]{0.8} \\
\hline

ET2CE (10 yr) & $\dots$ & $\dots$ & $\dots$ & 88.63\% \\
BDET2CE (1 yr) & $\dots$ & $\dots$ & $\dots$ & 78.21\% \\
BDET2CE (10 yr) & 91.48\% & 43.16\% & 32.30\% & 24.88\% \\

\hline\hline
\end{tabular}
\end{table}

\begin{table}
\centering
\caption{Relative $1\sigma$ uncertainties of the five AP model parameters, $C$, $D$, $f^{\mathrm{P}}$, $b_0$, and $b_1$, for $f^{\mathrm{P}}=0.4$ using 10 years of BDET2CE observations combined with CSST. Results are shown for different combinations of angular power spectra.}\label{tab:f_P_0.4}
\centering
\centering
\renewcommand{\arraystretch}{2}
\begin{tabular}{cccccc}
\hline\hline
\makebox[0.071\textwidth][c]{Type} & \makebox[0.071\textwidth][c]{$C$} &
\makebox[0.071\textwidth][c]{$D$} & \makebox[0.071\textwidth][c]{$f^{\mathrm{P}}$} & \makebox[0.071\textwidth][c]{$b_0$} & \makebox[0.071\textwidth][c]{$b_1$} \\
\hline

total & 5.35\% & 14.87\% & 43.16\% & 0.27\% & 0.32\% \\
gGW & 30.14\% & 81.77\% & 45.48\% & 26.60\% & 41.37\% \\
GWGW & 32.15\% & 90.53\% & 256.17\% & $\dots$ & $\dots$  \\
gg & $\dots$ & $\dots$ & $\dots$ & 0.27\% & 0.32\% \\

\hline\hline
\end{tabular}
\end{table}

\section{Conclusions}\label{sec:conclusions}

We have investigated how cross-correlating GW events with galaxies can identify a PBH contribution to the BBH merger population. The idea is that ABH and PBH mergers, residing in different host environments, trace the cosmic matter distribution with different clustering biases; measuring the GW clustering bias statistically against a galaxy survey therefore separates the two populations. Using ABH and early-PBH merger-rate models, we built mixed mock GW catalogs for the third-generation ET2CE network and the multiband BDET2CE network, combined them with a mock CSST photometric galaxy sample, and used a Fisher analysis of the galaxy autocorrelation, galaxy-GW cross-correlation, and GW autocorrelation spectra, including GW localization damping, to forecast the constraints.

Our central finding is that detecting a PBH contribution is considerably easier than measuring its fraction. Both detection diagnostics agree: the effective GW bias leaves the pure-ABH uncertainty band, and the model-distinguishing SNR rises above 4, once the PBH fraction reaches about 40\% for ET2CE and about 20\% for BDET2CE, each combined with CSST. Pinning down $f^{\mathrm{P}}$ as a free parameter is more demanding: even for BDET2CE the relative uncertainty is 91.48\% at $f^{\mathrm{P}}=0.2$, improving to 43.16\% at $f^{\mathrm{P}}=0.4$. The advantage of BDET2CE comes from its multiband sky localization, two to three orders of magnitude better than ET2CE. This pushes the localization-damping scale to higher multipoles and recovers the small-scale modes that carry most of the clustering information, most effectively at high redshift, precisely where the PBH-induced bias deviation is largest.

These results point to two complementary requirements for the method. A deep galaxy survey such as CSST is needed to reach the high redshifts where PBH mergers are relatively more abundant and the bias deviation is strongest, while accurate GW localization is needed to retain the small-scale clustering information that drives the sensitivity. The three-spectrum combination is equally important: the dense galaxy autocorrelation alone pins the galaxy bias to subpercent precision, breaking its degeneracy with the GW bias and tightening the entire parameter set. Together, these make GW-galaxy clustering a promising and largely independent route to the statistical identification of PBHs, one that the next generation of GW detectors, and especially multiband networks like BDET2CE, will bring within reach when paired with the wide, deep galaxy surveys now on the horizon.

\begin{acknowledgments}
This work was supported by the National Natural Science Foundation of China (Grants No. 12533001, No. 12575049, and No. 12473001), the National SKA Program of China (Grants No. 2022SKA0110200 and No. 2022SKA0110203), the China Manned Space Program (Grant No. CMS-CSST-2025-A02), and the National 111 Project (Grant No. B16009).
\end{acknowledgments}

\section*{Data Availability}
There are no publicly available research data or software supporting this manuscript. Requests for further information or data should be sent to 
the authors.

\bibliography{ref}

\end{document}